\documentclass[reprint,amsmath,amssymb, aps,]{revtex4-2}
\pdfoutput=1

\usepackage{braket}
\usepackage{graphicx}
\usepackage{amsmath}
\usepackage{hyperref}

\usepackage{xcolor}

\newcommand{\un}[1]{\mathrm{\:#1}}
\newcommand{\comment}[1]{}

\begin{document}
	
\title{Photon Number Resolving Detection with a Single-Photon Detector and Adaptive Storage Loop}
\date{November 21, 2023}
\author{Nicholas M. Sullivan}
\email{nm.sullivan@mail.utoronto.ca}
\affiliation{Department of Physics,	University of Ottawa, 25 Templeton Street, Ottawa, ON, Canada K1N 6N5}
\author{Boris Braverman}
\thanks{Present Address: Department of Physics, University of Toronto, 60 St. George Street, Toronto, ON, Canada, M5S 1A7}
\email{boris.braverman@utoronto.ca}
\affiliation{Department of Physics,	University of Ottawa, 25 Templeton Street, Ottawa, ON, Canada K1N 6N5}
\author{Jeremy Upham}
\affiliation{Department of Physics,	University of Ottawa, 25 Templeton Street, Ottawa, ON, Canada K1N 6N5}
\author{Robert W. Boyd}
\affiliation{Department of Physics,	University of Ottawa, 25 Templeton Street, Ottawa, ON, Canada K1N 6N5}
\affiliation{Department of Physics and Astronomy, University of Rochester, 275 Hutchison Road, Rochester, NY, 14627, USA}


\begin{abstract}
Photon number resolving (PNR) measurements are beneficial or even necessary for many applications in quantum optics. Unfortunately, PNR detectors are usually large, slow,  expensive, and difficult to operate. However, if the input signal is multiplexed, photon ``click'' detectors, that lack an intrinsic photon number resolving capability, can still be used to realize photon number resolution. Here, we investigate the operation of a single click detector, together with a storage line with tunable outcoupling. Using adaptive feedback to adjust the storage outcoupling rate, the dynamic range of the detector can in certain situations be extended by up to an order of magnitude relative to a purely passive setup. An adaptive approach can thus allow for photon number variance below the quantum shot noise limit under a wider range of conditions than using a passive multiplexing approach. This can enable applications in quantum enhanced metrology and quantum computing.
\end{abstract}

\maketitle

\section{Introduction}
\label{sec:intro}

Quantum optics and quantum photonics technologies are increasingly important for applications in domains such as sensing and computing. Many of these applications require the use of high-resolution and efficient photon detectors. In particular, the ability to count the number of incident photons on a detector is vital for applications including quantum-enhanced metrology \cite{demkowicz-dobrzanskiElusiveHeisenbergLimit2012a,motesLinearOpticalQuantum2015} and quantum computing \cite{zhongQuantumComputationalAdvantage2020,madsenQuantumComputationalAdvantage2022}. 

Important features of a photon number resolving (PNR) detector include detection efficiency and fidelity, dark count rate, speed, ease of operation, and dynamic range. Existing detectors lack the combination of dynamic range, detection fidelity and ease of operation that are relevant for many applications. Although single-photon detectors (SPDs) such as avalanche photodiodes are widely available and relatively robust, they are unable to distinguish one incident photon from a greater number, effectively providing a binary ``click'' signal.

Several PNR detector schemes have been developed to distinguish multi-photon events. Transition edge sensor (TES) detectors \cite{gerritsExtendingSinglephotonOptimized2012,litaCountingNearinfraredSinglephotons2008} and microwave kinetic inductance detectors \cite{guoCountingInfraredPhotons2017} have demonstrated PNR capabilities up to tens of photons with low noise, although they are limited by their low speed and very low required operation temperature. Superconducting nanowire single photon detectors (SNSPD) \cite{zhuResolvingPhotonNumbers2020,schmidtCharacterizationPhotonNumberResolving2019} have high detection efficiencies and fast count rates, but tend to saturate at low photon numbers. CMOS detectors \cite{maMmPitchQuantaImage2016} and intensified charge coupled devices (ICCDs) \cite{ORCAQuestQCMOSCamera} have also demonstrated some PNR capability at low photon numbers, but either have low detection efficiency or slow readout. 

A PNR detector can alternatively be constructed by multiplexing the input, so that the photons can be counted with multiple detection events from one or more SPDs. In this type of architecture, photons can be treated as classical particles since interference does not play an important role. Each detector or time slot can thus be treated as a bin that receives photons with a certain probability. There are several ways to characterize PNR detectors using information theory \cite{enkPhotodetectorFiguresMerit2017} and detector tomography \cite{humphreysTomographyPhotonnumberResolving2015,feitoMeasuringMeasurementTheory2009a}, and previous studies have compared the merits of various multiplexing-based approaches \cite{jonssonEvaluatingPerformancePhotonnumberresolving2019, schapelerInformationExtractionPhotoncounting2022a}. 

In some approaches, the input signal is spatially multiplexed to an array of detectors. This can be achieved using beam splitters \cite{heilmannHarnessingClickDetectors2016,hlousekAccurateDetectionArbitrary2019,teoProspectsMultiportDevices2019} or flood illumination \cite{jiangPhotonnumberresolvingDetector102007}, while using fast optical switches can reduce detector deadtime \cite{liuSwitchableDetectorArray2019,castellettoReducedDeadtimeHigher2007}. These spatially-multiplexing methods have been evaluated in detail \cite{jonssonPhotoncountingDistributionArrays2020}, and explicit expressions for the detection probabilities have been derived \cite{miattoExplicitFormulasPhoton2018}. 

There are also several methods to temporally multiplex the input signal, either by using $N$ beam splitters and delay lines to produce a detector with $2^N$ equal-fraction time bins \cite{achillesFiberassistedDetectionPhoton2003, natarajanQuantumDetectorTomography2013,jonssonTemporalArraySuperconducting2020}, or using a loop-based detector consisting of a storage loop and tunable coupling $\epsilon$ to multiplex the photons into an unlimited number of decreasing-probability time bins \cite{banaszekPhotonCountingLoop2003,rehacekMultiplephotonResolvingFiberloop2003,tiedauHighDynamicRange2019,webbPhotostatisticsReconstructionLoop2009}. 

\begin{figure}[btp] 
    \includegraphics[width=1.0\linewidth]{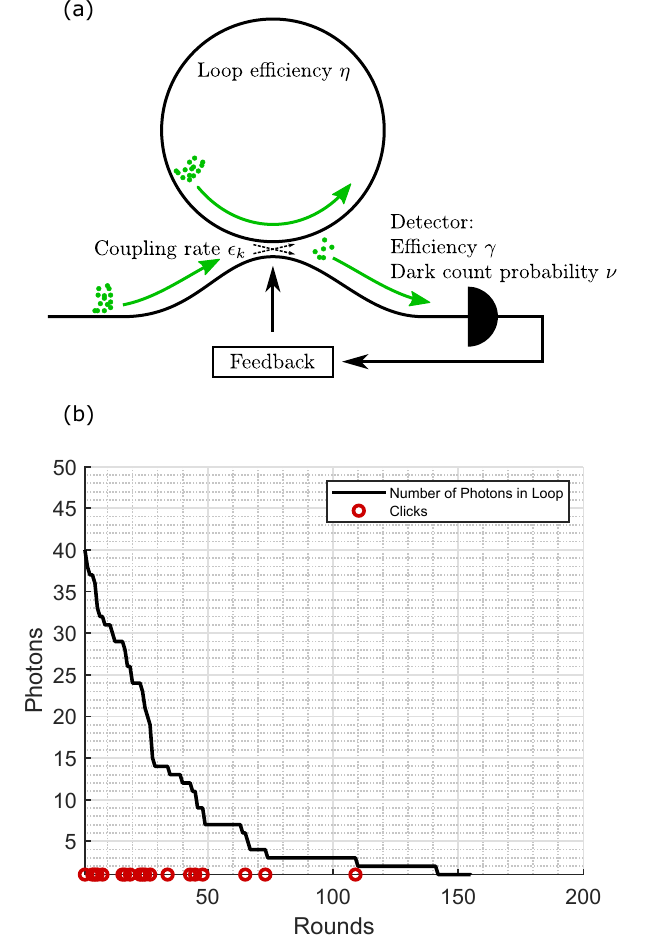}
	\caption{(a) A loop-based photon number resolving setup allows a pulse of photons to be stored and sent to a single-photon detector in multiple rounds. Here, we investigate the improvements that can be offered by dynamic feedback on the outcoupling rate $\epsilon_k$, taking into account the loop efficiency $\eta$, and detector characteristics ($\gamma$ and $\nu$) (b) When the outcoupling rate is kept constant, most detections occur in the early rounds, many of which correspond to multiple-photon events. As a result, many photons are not detected. In the example in the figure, $\epsilon = 0.02$. The initial number of photons is $40$, but only $19$ clicks are recorded by the detector.}
    \label{fig:Diagram}
\end{figure}

The loop-based PNR setup is distinct from the other multiplexing-based methods in that the photons can be stored and released in a controlled way while detection clicks are recorded. This setup is constructed of standard components: SPDs, low-loss fiber, and tunable fiber couplers. First, an input signal with $N_0$ photons is coupled from the input line into a fiber delay loop which has photon storage efficiency $\eta$ (Figure \ref{fig:Diagram}). After this delay, the pulse returns to the fiber coupler, which has been tuned from full coupling ($\epsilon \approx 1$) to a much smaller coupling ($\epsilon \ll 1$). Here, a fraction of the photons in the loop are sent to an SPD (having efficiency $\gamma$ and dark count rate $\nu$), which produces a binary detection result $d_1 \in \{0,1\}$. The majority of the photons remain in the fiber, and the cycle repeats, with the outcoupling ratio possibly tuned after each round, which we denote by $\epsilon_k$ in round $k$. Table \ref{table:symbols1} of Appendix \hyperref[sec:appendixB]{B} has a complete list of the variables used in this paper. When all photons have been detected or lost after some number of rounds $k$, the experimenter is left with a series of detection results $\vec{d}_k := (d_1,d_2,...,d_k)$, from which one can estimate the initial number of photons $N_0$. 

In this work, we introduce an adaptive method of PNR detection in the loop architecture that demonstrates improved accuracy, speed, and dynamic range. Every approach previously taken has the limitation of being unable to adapt to the number of photons within a single measurement, due to their static architectures. In particular, a passive multiplexing detector suited for discriminating small photon numbers will become over-saturated at large photon numbers, while a detector suited for large photon numbers will incur higher losses than necessary at small photon numbers. Here, we demonstrate a method of using the real-time loop detection results $\vec{d}_k$ as feedback for tweaking the outcoupling rate $\epsilon_k$, and evaluate the performance of this approach using Monte Carlo simulations. This approach produces a photon-number estimate that matches or improves on the estimate of any passive approach, while operating over a wide dynamic range and requiring fewer detection rounds. Comparable single-shot adaptive detection has been demonstrated for homodyne detection of optical phase \cite{armenAdaptiveHomodyneMeasurement2002a}, but not yet for photon number. 

This work is divided in several sections. In \hyperref[sec:mathback]{Mathematical Background}, we discuss Bayes' Theorem and how it applies to our loop-based PNR setup. There, we also introduce the method we use to adaptively tune the outcoupling rate $\epsilon_k$. In \hyperref[sec:methods]{Methods}, we discuss how the simulations were performed, and introduce the metrics we used to evaluate the performance of the passive and adaptive approaches under different conditions. In \hyperref[sec:results]{Results}, we compare the merits of the passive and adaptive approaches, with respect to the earlier introduced metrics. Finally, in \hyperref[sec:discussion]{Discussion}, we consider the performance of the adaptive approach, and possible ways this analysis can be extended to other loop-based setups. Appendices summarizing the calculation of the Bayesian updating probability (Appendix \hyperref[sec:appendixA]{A}) and the list of variables and symbols used in this paper (Appendix \hyperref[sec:appendixB]{B}) are also included.

\section{Mathematical Background}
\label{sec:mathback}

The purpose of this section is twofold. First, we introduce the relevant mathematical background necessary for performing the Bayesian estimate of the initial photon number, based on the detection results. Then, we motivate the method we use to adaptively choose the outcoupling rate $\epsilon_k$, based on the principle of maximizing the expected ratio of information gained to information lost. All variables and distributions introduced in this section are enumerated in Table \ref{table:symbols2} of Appendix \hyperref[sec:appendixB]{B}.

The first question we want to consider is: How does one obtain an estimate of the initial photon number from the detection results? Clearly, the estimate $N_{est}$ is a function of the detection results $\vec{d}_k$, in addition to being implicitly dependent on the system parameters (loop efficiency $\eta$, outcoupling rate $\epsilon$, etc.). Thus, our goal is to find a suitable $N_{est} (\vec{d}_k)$, which we will accomplish using Bayes' Theorem.

Bayes' Theorem states that for any events $A$ and $B$, the conditional probabilities $P(A|B)$ and $P(B|A)$ are related:

\begin{equation}
    \begin{split}
        P(A|B) & = \frac{P(B|A)P(A)}{P(B)}.
    \end{split}
\end{equation}

This provides an equation that describes how beliefs (about event $A$) should be updated as more information becomes known (event $B$). In particular, Bayes' Theorem takes a prior probability distribution $P(A)$ and produces a posterior probability distribution $P(A|B)$, requiring only a conditional probability $P(B|A)$. In our case, we would like to find the probability that there were $N_0$ initial photons, given the detection record $\vec{d}_k$. However, it turns out that it is even more useful to also find the probability of $N_k$ photons being in the loop during round $k$, in addition to $N_0$ initial photons. This probability can thus be found as follows:

\begin{equation}
    \begin{split}
        P(N_k,N_0|\vec{d}_k) & = \frac{P(N_k,\vec{d}_k|N_0)P(N_0)}{P(\vec{d}_k)}.
    \end{split}
    \label{eq:posterior}
\end{equation}

Note that $P(N_k,\vec{d}_k|N_0)$ is just the probability of recording the given detection results $\vec{d}_k$, and having $N_k$ photons remaining in the loop, given $N_0$ initial photons. As a probability distribution over all detection histories, this is difficult to determine. But since we are only interested in a probability distribution over $N_k$ for a particular detection history, this is feasible to do.  

As noted in the introduction, the photons in this setup can be modeled simply as a collection of classical particles in the loop. As a result, the probability of a detection in any given round is determined solely by the current number of photons in the loop. Thus, given some number of photons $N_{k-1}$ in the loop during round $k-1$, we can find the probability of $N_{k}$ photons remaining in the loop, and a click either being registered or not (which we represent by $d_k$, taking on values $0$ or $1$). This we can denote as $P(N_{k}, d_{k}| N_{k-1})$. This probability implicitly depends on the parameters of the setup, including $\epsilon_k$, $\eta$, $\gamma$ and $\nu$. Thus, the easiest way to calculate the probability $P(N_k,\vec{d}_k|N_0)$ is in a recursive manner, updating as follows:

\begin{equation}
    \begin{split}
        P(N_k, \vec{d}_k|N_0) & = \sum_{N_{k-1}} P(N_k, d_k|N_{k-1}) P(N_{k-1}, \vec{d}_{k-1}|N_0).
    \end{split}
    \label{eq:update}
\end{equation}

The term $P(N_{k}, d_{k}| N_{k-1})$, described in the previous paragraph, can be calculated from the characteristics of the setup (see Appendix \hyperref[sec:appendixA]{A}), which include the loop efficiency $\eta$, outcoupling rate $\epsilon$, detector efficiency $\gamma$ and detector dark count probability (during delay time $\tau$) $\nu$. This is given by:

\begin{widetext}
\begin{equation}
    \begin{split}
        P(N_{k}, d_{k}| N_{k-1}) & = 
        \begin{cases}
            \rho(N_k; N_{k-1}, \eta, \epsilon_k, \gamma, \nu) & d_{k} = 0\\
            B(N_k; N_{k-1}, \eta(1-\epsilon_k)) - \rho(N_k; N_{k-1}, \eta, \epsilon_k, \gamma, \nu) & d_{k} = 1\\
        \end{cases},
    \end{split}
\end{equation}
where
\begin{equation}
    \begin{split}
        \rho(N_k; N_{k-1}, \eta, \epsilon_k, \gamma, \nu) & = (1-\nu)(1 - \eta\epsilon_k\gamma)^{N_{k-1}} B\left(N_k; N_{k-1}, \frac{\eta(1-\epsilon_k)}{1 - \eta\epsilon_k\gamma}\right),
    \end{split}
\end{equation}
\end{widetext}
and $B(k; n, p)$ is the binomial distribution function:
\begin{equation}
    \begin{split}
        B(k; n, p) & = \binom{n}{k} p^k (1-p)^{n-k}.
    \end{split}
\end{equation}

By restricting our attention to one particular detection sequence $\vec{d}_k$, we can represent the probability $P(N_k, \vec{d}_k|N_0)$ as a $(N_{max} + 1)\times(N_{max} + 1)$ matrix $\mathbf{P} (\vec{d}_k)$, where $N_{max}$ is a cap on the number of incident photons we consider. The update probability $P(N_k,d_k|N_{k-1})$ can similarly be represented as a matrix $\mathbf{R} (d_{k})$ (depending implicitly on $\epsilon_k$), whose elements are given by:
\begin{equation}
    \begin{split}
        [\mathbf{P} (\vec{d}_k)]_{m,n} & = P(m, \vec{d}_k|n)\\
        [\mathbf{R} (d_{k})]_{m,n} & = P(m, d_{k}|n).
    \end{split}
    \label{eq:ProbMat}
\end{equation}

Consequently, we can translate the Bayesian updating process into matrix multiplication, where the matrix $\mathbf{R}$ is chosen based on whether a click was recorded during that round.
\begin{equation}
    \begin{split}
        \mathbf{P} (\vec{d}_{k}) & = \mathbf{R} (d_{k}) \mathbf{P} (\vec{d}_{k-1}).
    \end{split}
\end{equation}

\begin{figure}[btp] 
    \centering
    \includegraphics[width=1.0\linewidth]{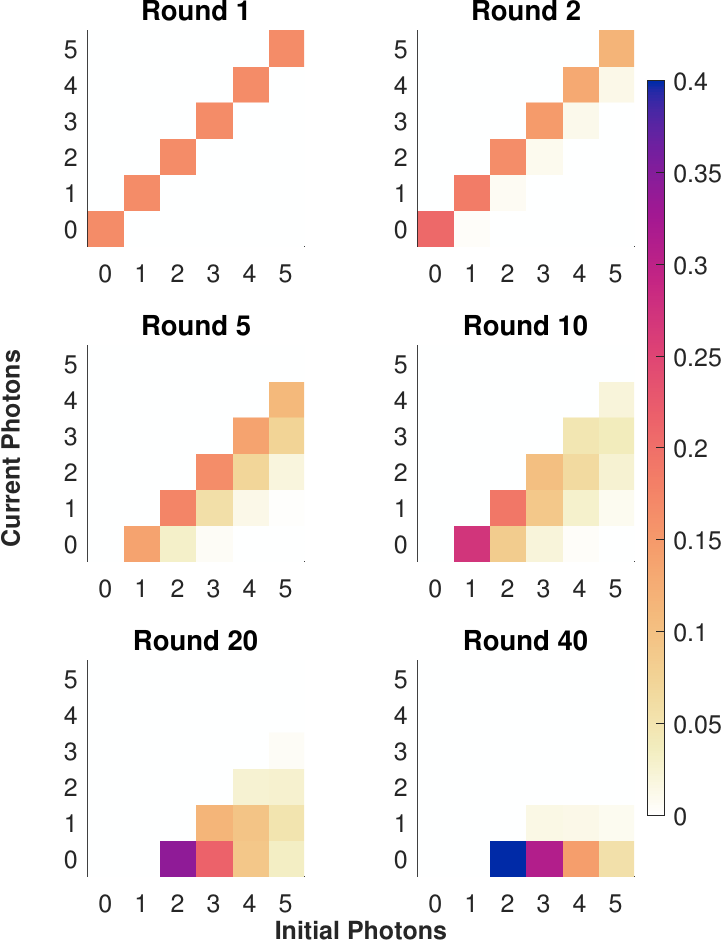}
	\caption{Before each detection round $k$, the state of knowledge can be encoded in a posterior probability distribution for the number of photons in the loop, both initially ($N_0$) and in the current round ($N_k$). These matrices can be used to determine the final posterior distribution for $N_0$, and in an adaptive approach, to inform a good value for the loop outcoupling rate $\epsilon_k$. In these visualizations, we observe an evolution of the posterior probability matrix $\mathbf{P} (\vec{d}_{k})$ for a single realization of $\vec{d}_k$, where detections were recorded in rounds $k = 2$ and $k = 15$. Starting from a uniform distribution in Round $1$, $\mathbf{P} (\vec{d}_{k})$ is updated in each round using Bayes' theorem, resulting in a distribution close to the true value ($N_0 = 3$) in Round $40$. The simulation parameters are $\eta = 0.99$, $\gamma = 0.9$, $\nu = 1\times 10^{-6}$, and $\epsilon = 0.1$.}
    \label{fig:BayesianProbMatrix}
\end{figure}

After $k = 0$ rounds, no photons can have been lost, and thus $P(N, \vec{d}_{0} | N_0)$ is only non-zero when $N = N_0$. Assuming a uniform prior on the number of photons in the loop, we know that the conditional probability $\mathbf{P} (\vec{d}_{0})$ is equal to the identity matrix (where $\vec{d}_{0}$ is the empty set). Thus, by recursively iterating this process, we find that $\mathbf{P} (\vec{d}_{k})$ can be represented:
\begin{equation}
    \begin{split}
        \mathbf{P} (\vec{d}_{k}) & = \left[\prod_{j=1}^{k} \mathbf{R} (d_{j}) \right] \mathbf{P} (\vec{d}_{0})\\
        & = \prod_{j=1}^{k} \mathbf{R} (d_{j})\\
    \end{split}
\end{equation}

Note that this reformulation of probabilities in matrix form does not allow us to sample all possible histories $\vec{d}_k$, but can be used to speed Monte Carlo sampling of the detector histories by taking advantage of matrix multiplication libraries. Additional computational efficiency can be gained by precomputing the matrices $\mathbf{R} (d)$, especially if the loop coupling parameter is kept constant. 

Now that we know how to update the conditional probability $P(N_k,\vec{d}_k|N_0)$, only the prior probability $P(N_0)$ is necessary to calculate the posterior probability $P(N_k, N_0|\vec{d}_k)$, as seen in Eq. \eqref{eq:posterior}. The probability $P(\vec{d}_k)$ in the denominator, containing neither $N_0$ nor $N_k$, merely becomes a normalization factor for the posterior distribution. This posterior distribution is used later for the purposes of quantifying the information gain. Further, the posterior probability $P(N_0|\vec{d}_k)$ for the initial number of photons can be easily computed by summing $P(N_k,N_0|\vec{d}_k)$ over $N_k$. If $k$ is large enough that there are likely no photons remaining in the loop, we know that no more clicks are likely to be observed, and the experiment can be ended. One can then produce an estimate $N_{est}$ from $P(N_0|\vec{d}_n)$, either by choosing the value of $N_0$ for which $P(N_0|\vec{d}_n)$ is maximized (known as the Maximum Likelihood Estimate, or MLE), or by choosing the mean value of $N_0$ according to the distribution $P(N_0|\vec{d}_n)$. 

The benefit of using such a posterior probability (or conditional probability) matrix is that it represents the total information one can gain from the data at a particular time. The higher the correlation between $N_0$ and $N_k$, the more that can be learned by knowing the number of photons currently in the loop. The matrix also contains information about the number of photons currently in the loop, which allows the outcoupling rate to be tuned to learn the most information overall.

Figure \ref{fig:BayesianProbMatrix} visually displays an example posterior probability matrix $P(N_k, N_0|\vec{d}_n)$ during different rounds, for $N_{max} = 5$, $N_0 = 3$. Initially, the posterior probability matrix is specified according to a uniform prior probability distribution $P(N_0) = \frac{1}{6}$, with all off-diagonal terms identically zero (since $k = 0$). In general, we always know that $N_k \leq N_0$, and as a result, the matrix is always upper-triangular. In the subsequent rounds, the posterior probability matrix updates based on whether a detection has been made. After $40$ rounds, the number of photons estimated to be in the loop drops close to $0$, and the distribution $P(N_0|\vec{d}_n)$ can be taken as the final estimate. Two detector clicks were recorded in total (in rounds $k = 2$ and $k = 15$), and thus the final estimate (sub-figure for Round 40 in the bottom row of Figure \ref{fig:BayesianProbMatrix}) finds $N_0 = 2$ to be the most likely number of initial photons, followed by $N_0 = 3$ and $N_0 = 4$. The MLE of this example would thus be $N_{MLE} = 2$, while the mean value estimate is slightly greater ($N_{est} \approx 2.8$). At this point, it is possible that there is still a photon in the storage loop, but it is more likely that one was lost due to loop loss or detector inefficiency. 

The second question we must consider is how one can optimally choose the outcoupling rate $\epsilon_k$ for each round. This is more difficult to address, and depends on the capabilities of the setup. We can consider a passive setup, in which the outcoupling rates are set before the measurement. On the other hand, we can also consider an adaptive setup, in which the outcoupling rates are changed during the measurement based on the detector results.

In either case, the ultimate goal of choosing the outcoupling rates is to maximize the expected information gained about the photon number by the end of the experiment. If we only cared about extracting the maximum information possible from a single detection round, we would adjust the setup such that our expected variance in the distribution of $N_0$ is minimized after that particular round. Depending on the click result, the prior distribution would be partitioned into one of two distributions. In the ideal case, one bit of information about the distribution would be gained, and thus each scenario would have equal probability. One possible strategy, therefore, is to adjust $\epsilon$ each round such that the click probability is $50\%$.

However, this strategy does not work well in practice, for several reasons. If the per-round click probability is made to be as high as $50\%$, then many of those clicks will consist of more than one photon hitting the detector. Since our detector cannot distinguish between one or more photons, these extra photons are lost, and contribute to a higher variance in the posterior distribution. Furthermore, because we have multiple rounds to detect photons, we need to account for the photons lost within the loop, which could have been used to gain more information at a later point in the experiment. Therefore, we must consider not only the information gained during any given round, but how that choice would affect the information that can be gained in later rounds. 

One way to quantify the information gain is through the Kullback-Leibler (KL) divergence of the posterior distribution of $N_0$ against the priors we assign to $N_0$ \cite{kullbackInformationSufficiency1951}:

\begin{equation}
    \begin{split}
        I_{G,k} & = D_{KL}(P(N_0|\vec{d}_k)||P(N_0))\\
        & = \sum_{N_0}P(N_0|\vec{d}_k)\log\left(\frac{P(N_0|\vec{d}_k)}{P(N_0)}\right).
    \end{split}
\end{equation}

Although the KL divergence is more properly used as a way of quantifying the divergence between the true probability distribution and a model, it is appropriate here because the distribution $P(N_0|\vec{d}_k)$ is usually much closer to the true distribution than $P(N_0)$. Similarly, one way to quantify the loss of information through inefficiencies is due to the change in information about $N_0$ available to learn based on the number of photons remaining in the loop. Specifically, we quantify the available information as the expected value of the KL divergence between $P(N_0|N_k,\vec{d}_k)$ and the initial distribution of $N_0$:

\begin{equation}
    \begin{split}
        I_{A,k} & = \sum_{N_k}P(N_k|\vec{d}_k)D_{KL}(P(N_0|N_k,\vec{d}_k)||P(N_0))\\
        & = \sum_{N_k,N_0}P(N_k|\vec{d}_k)P(N_0|N_k,\vec{d}_k)\log\left(\frac{P(N_0|N_k,\vec{d}_k)}{P(N_0)}\right)\\
        & = \sum_{N_k,N_0}P(N_k,N_0|\vec{d}_k)\log\left(\frac{P(N_k,N_0|\vec{d}_k)}{P(N_k|\vec{d}_k)P(N_0)}\right).
    \end{split}
\end{equation}

The available information thus represents maximum information that one could possibly learn about $N_0$ after a certain number of rounds. The expected value of $I_{A,0}$ (available information before the start of the experiment) is simply the Shannon entropy of the prior distribution $P(N_0)$, while $I_{G,0}$, (the information gained), will be precisely 0, because no detector result will have been recorded. Assuming we allow the experiment to run long enough that the photon survival probability is very small, then at the end of the experiment, $N_k=0$ with 100\% probability, so the available information and the information gain will be equal. The goal of choosing the outcoupling rates is for this value ($I_{G,R} = I_{A,R}$) to be as high as possible by the end of the experiment, where $R$ is the total number of rounds. 

In the passive case, one must have a good idea of $N_0$ initially, to properly choose the outcoupling rates. If the outcoupling rates are too small, most of the photons will be lost to the intrinsic loss within the loop, and not as many photons will be detected by the SPD, ultimately decreasing the overall efficiency of this approach. With many photons lost in the loop, the available information is correspondingly degraded because we don't know exactly how many photons were lost. On the other hand, if the outcoupling rates are too large, then the detector will be saturated on most rounds, also resulting in undetected photons. Only for a small range of $N_0$ will the photon number estimate be optimal relative to the fixed imperfections of the apparatus. 

In the adaptive case, the requirement of our initial knowledge of $N_0$ is somewhat relaxed. This is because, if a large fraction of the detection rounds produce clicks, the outcoupling rate can be reduced to store the photons for longer in the loop, while if a large fraction of the detections do not produce clicks, the outcoupling rate can be increased to send more photons to the detector. 

To properly choose the outcoupling rates in the adaptive setup is a subtle matter. One must first decide on a prior probability distribution $P(N_0)$, and update the posterior probability $P(N_k, N_0|\vec{d}_k)$ after each round $k$. Then, based on this posterior probability, one must choose the appropriate value of $\epsilon$ to maximize the expected information gain in the next round, taking into account the information lost in the loop. Since it is inevitable that some information is lost, we would like to optimize the ratio of the information gained to the information lost, so that we gain the most information possible during the entire experiment. Thus, in round $k$ we would like to choose $\epsilon_k$ such that:

\begin{equation}
    \begin{split}
        \Bigg|\frac{\langle I_{G,k}\rangle - I_{G,k-1}}{\langle I_{A,k}\rangle - I_{A,k-1}}\Bigg|
    \end{split}
    \label{eq:InfoRatio}
\end{equation}

is maximized for our choice, where $\langle\cdot\rangle$ here denotes the expected value, given current knowledge. 

\section{Methods}
\label{sec:methods}

In this section, we describe the details of how the simulations were performed, based on the mathematical background introduced in the \hyperref[sec:mathback]{Mathematical Background} section. Additionally, we introduce and discuss the metrics we use to evaluate and compare the performance of the passive and adaptive methods. All variables and symbols introduced in this section are enumerated in Table \ref{table:symbols3} of Appendix \hyperref[sec:appendixB]{B}.

As described in the \hyperref[sec:intro]{Introduction}, the loop-based PNR detector we simulated consists of a fiber loop with efficiency $\eta$, and a detector with efficiency $\gamma$ and dark count rate $\nu$. Additionally, the outcoupling rate $\epsilon$ is allowed to vary between rounds, so that $\epsilon = \epsilon_k$ during round $k$. First, we fix the initial number of photons $N_0$ that are input into the setup, and an initial probability distribution $P(N_0)$ is chosen, reflecting our belief of the input photon distribution. For all of our simulations, we use a uniform distribution from $0$ to $N_{max}$ photons. Next, for each round, the probability of a click result with some number of remaining photons in the loop, $P(N_{j+1}, d_{j+1}; N_{j})$, is calculated from \eqref{eq:ProbMat}. A result is chosen with probability given by this distribution, and the click result is then used to determine the updated Bayesian probability distribution $P(N_{j+1}, N_0)$ from $P(N_{j}, N_0)$. The experiment is continued until the estimated number of photons remaining in the loop is below some threshold $N_t$ (for example, $N_t = 0.5$). If we assume that loop inefficiency is the dominant form of loss, then we can assume this is true after approximately $\frac{\ln (N_{max}/N_t)}{1-\eta}$ rounds. Alternatively, the current belief matrix can be used to determine when the estimated number of photons remaining is below $N_t$. The final distribution gives our estimate of the initial number of photons that have been fed into the setup.

The most important parameters to vary are the initial number of photons $N_0$, the loop efficiency $\eta$ and the detector efficiency $\gamma$. The typical values chosen for these physical parameters in the simulation are motivated by the performance specifications of real-world devices. The performance of the estimator over different initial photon numbers determines the dynamic range of the setup, while the loop efficiency and detector efficiency both have important effects on the estimate uncertainty and error. The number of useful rounds and the best outcoupling rate are also strongly determined by loop efficiency and detector efficiency, while the dark count probability $\nu$ is generally small enough to not have a large effect.

During each round, we are free to adjust the value of $\epsilon$ according to some strategy or algorithm decided beforehand. We consider both a passive approach and an adaptive approach, as discussed in the \hyperref[sec:intro]{Introduction} and \hyperref[sec:mathback]{Mathematical Background} sections. In the passive approach, the outcoupling rate $\epsilon$ is kept constant and chosen before the simulation, so each Bayesian update matrix $\mathbf{R}$ will be identical and can be precomputed, which has the benefit of speeding up the computations. For the adaptive approach, the value of $\epsilon_k$ is chosen in real time using the analysis of the detector results. In this case, the current belief distribution must be used to estimate the number of photons in the loop. Then, $\epsilon_k$ can be chosen in such a way to maximize the ratio between the expected information gained and lost due to this choice (see Eq. \eqref{eq:InfoRatio} in the \hyperref[sec:mathback]{Mathematical Background} section), in order to maximize the total information gained about the initial photon number in the course of the entire experiment. 

We end the simulation with the click record, the Bayesian belief matrix, and the actual number of photons in the input. Additionally, the history of the Bayesian belief matrix can be recorded, if desired. By repeating this simulation over many trials ($N_{trials} = 1000$), we can determine the performance of each strategy for choosing the outcoupling rate. The quantities of note for each specific trial are the estimated initial number of photons $N_{est} = \sum_{N_0} N_0 P(N_0)$, the estimated variance from Bayesian analysis, $\mathrm{Var}_{est} = \sum_{N_0} (N_0-N_{est})^2 P(N_0)$, and the maximum likelihood estimate (MLE) $N_{MLE} = \arg\max P(N_0)$.

Next, by analyzing many trials, we can determine the mean photon estimate $\langle N_{est} \rangle$, the mean estimated variance $\langle \mathrm{Var}_{est} \rangle$, the variance of estimates $\mathrm{Var}(N_{est}) = \langle (N_{est} - \langle N_{est} \rangle)^2 \rangle$, and the mean square error (MSE) in the photon estimate $\langle (N_{est} - N_0)^2 \rangle$, where $\langle \cdot\rangle$ denotes the average over all trials. Since the random trials are a representative sample of the space of possible outcomes, these sample means closely approximate a proper average over all possible outcomes. If the estimator is unbiased, the statement $\langle N_{est} \rangle = N_0$ must be satisfied for all $N_0$. In this case, the MSE would be equal to the variance of estimates $\mathrm{Var}(N_{est})$. However, for a biased estimator the MSE will always be larger than $\mathrm{Var}(N_{est})$, and will be a better measure of the estimate accuracy. 

The primary quantities we use to compare the approaches are the MSE, as well as $R$, the mean number of rounds needed until the estimated number of photons remaining in the loop is below the threshold $N_t = 0.5$. These quantities serve to measure the accuracy and repetition rate at which the detection process can be performed. These serve to compare the adaptive and passive approaches when the experimental parameters ($\eta$, $\gamma$, and $\nu$) and the initial photon number $N_0$ are fixed. 

It is helpful to consider the optimal performance that a loop-based detector could theoretically achieve, given a loop efficiency $\eta$ and detector quantum efficiency $\gamma$. If we only consider detector efficiency and loop efficiency, the total efficiency associated with the $k$th round detector bin is $\gamma\eta^k$. That is, if one photon is sent into the storage loop with no outcoupling, and the loop outcoupling switched on completely after the $k$th round, then the total probability of this photon being detected is $\langle n_k \rangle = \gamma\eta^k$. According to binomial statistics, the variance of this detection result will be $\mathrm{Var}(n_k) = \gamma\eta^k(1 - \gamma\eta^k)$. 

The optimal loop-based detector for $N_0$ incident photons can thus be modelled by a collection of detectors with efficiencies $\gamma\eta^k$, where exactly one photon is sent to each of the first $N_0$ detectors. Clearly, this cannot occur in a real loop-based detector where multi-photon events are unavoidable, but this can nevertheless be used to gauge the performance of the adaptive and passive approach estimates. Given $N_0$ incident photons, the number of clicks $n$ will have mean and variance as given below.

\begin{equation}
    \begin{split}
        \langle n \rangle & = \sum_{k = 0}^{N_0-1} \gamma \eta^k\\
        & = \gamma\frac{1 - \eta^{N_0}}{1 - \eta}\\
        \mathrm{Var}(n) & = \sum_{k = 0}^{N_0-1} \gamma \eta^k(1 - \gamma \eta^k)\\
        & = \gamma\frac{1 - \eta^{N_0}}{1 - \eta} - \gamma^2\frac{1 - \eta^{2N_0}}{1 - \eta^2}.
    \end{split}
\end{equation}

Suppose we use an estimator $N_{est}$ that is unbiased for any input photon number $N_0$. Then $\langle N_{est} \rangle = N_0$. We can approximate the variance in this estimator by simply scaling the variance in the number of clicks by $\left(\frac{N_0}{\langle n \rangle}\right)^2$. 

\begin{equation}
    \begin{split}
        \mathrm{Var}(N_{est}) & \approx \left(\frac{N_0}{\langle n \rangle}\right)^2\mathrm{Var}(n)\\
        & = N_0^2\left(\frac{\gamma(1 - \eta)}{1 - \eta^{N_0}}\right)^2\\
        & \times \left(\gamma\frac{1 - \eta^{N_0}}{1 - \eta} - \gamma^2\frac{1 - \eta^{2N_0}}{1 - \eta^2}\right)\\
        & = N_0^2\frac{1 - \eta}{1 - \eta^{N_0}}\left(\frac{1}{\gamma} - \frac{1 + \eta^{N_0}}{1 + \eta}\right).
    \end{split}
\end{equation}

For $N_0(1 - \eta) \ll 1$, this has the form $\mathrm{Var}(N_{est}) \approx N_0\left(\frac{1}{\gamma} - 1\right)$, while for $N_0(1 - \eta) \gg 1$, this can be approximated by $\mathrm{Var}(N_{est}) \approx N_0^2(1 - \eta)\left(\frac{1}{\gamma} - \frac{1}{1 + \eta}\right)$. This conforms to the intuition that the detector efficiency is more important for small numbers of photons, while loop efficiency has greater importance for large photon numbers. Although this performance is clearly not possible in a real experiment, it represents a limit on the resolution possible with loop efficiency $\eta$ and detector efficiency $\gamma$, and a benchmark for the adaptive strategy. 

\section{Results}
\label{sec:results}

\begin{figure}[btp] 
    \includegraphics[width=1.0\linewidth]{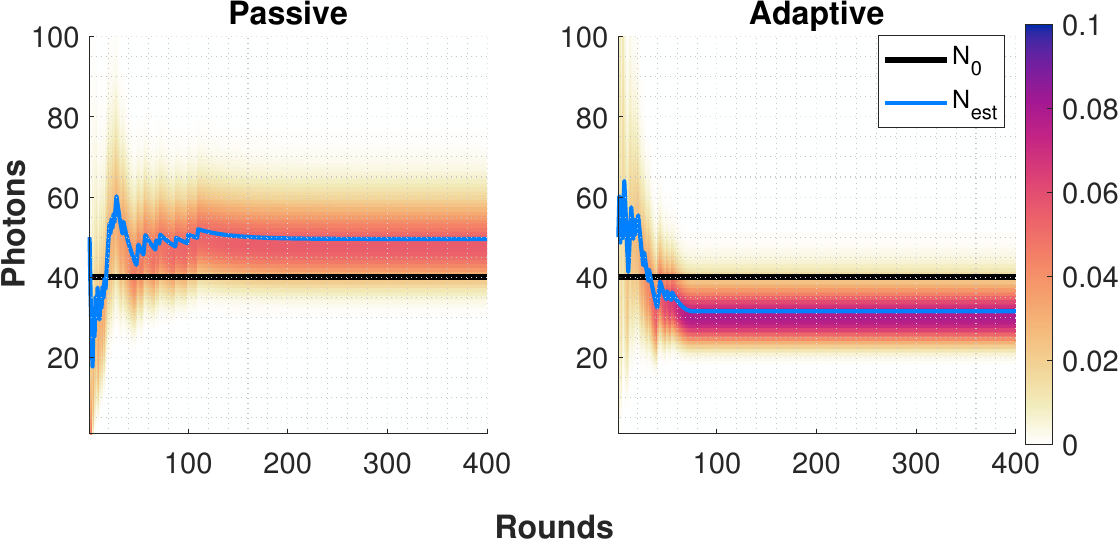}
	\caption{Sample evolutions of the initial photon number estimate using the passive (constant outcoupling, $\epsilon = 0.02$) and adaptive strategies illustrates that the adaptive strategy is capable of producing a more precise estimate in fewer rounds. The mean estimate (blue line) is displayed overtop with a plot of the posterior probability $P(N_0|\vec{d}_k)$. Due to stochastic effects in the detector, estimates from a single detection such as these are not always accurate, but are unbiased on average over many trials. The simulation had initial photon number $N_0 = 40$, and parameters $\gamma = 0.9$, $\eta = 0.99$ and $\nu = 1.0\times 10^{-6}$. With these parameters, the passive approach has bias $\langle N_{est} - N_0 \rangle \approx 1.7$ and MS error $\langle (N_{est} - N_0)^2 \rangle \approx 43$, while the adaptive approach has bias $\langle N_{est} - N_0 \rangle \approx 1.4$ and MS error $\langle (N_{est} - N_0)^2 \rangle \approx 35$.  }
    \label{fig:EstimatePassiveAdaptive}
\end{figure}

Below, we compare the merits of the adaptive and passive methods for various outcoupling rates, both during individual trials and in aggregate performance. For reference, we first show a plot of the estimates of $N_0$ associated with a representative experimental trial. On the left side of Figure \ref{fig:EstimatePassiveAdaptive}, we show how the estimated number of initial photons (using the mean estimate) evolve for a passive trial with $N_0 = 40$ initial photons. For comparison purposes, we choose the best-performing passive approach for these experimental conditions, with outcoupling rate $\epsilon = 0.02$, as discussed later. This is overlaid on a heatmap plot of the posterior $P(N_0 | \vec{d}_k)$. Initially, the estimate of $N_0$ varies quite quickly as many photon detection events are recorded, but changes become less frequent as the trial progresses and fewer photons remain in the loop. As expected, the estimated initial photon count eventually settles to a value near $N_0$. In estimates for the number of initial photons, discontinuities occur in tandem with photon detection events. The estimate of $N_0$ tends to decay in the rounds when no clicks occur, and jump noticeably whenever a click does occur. 

\begin{figure*}[btp] 
    \includegraphics[width=1.0\linewidth]{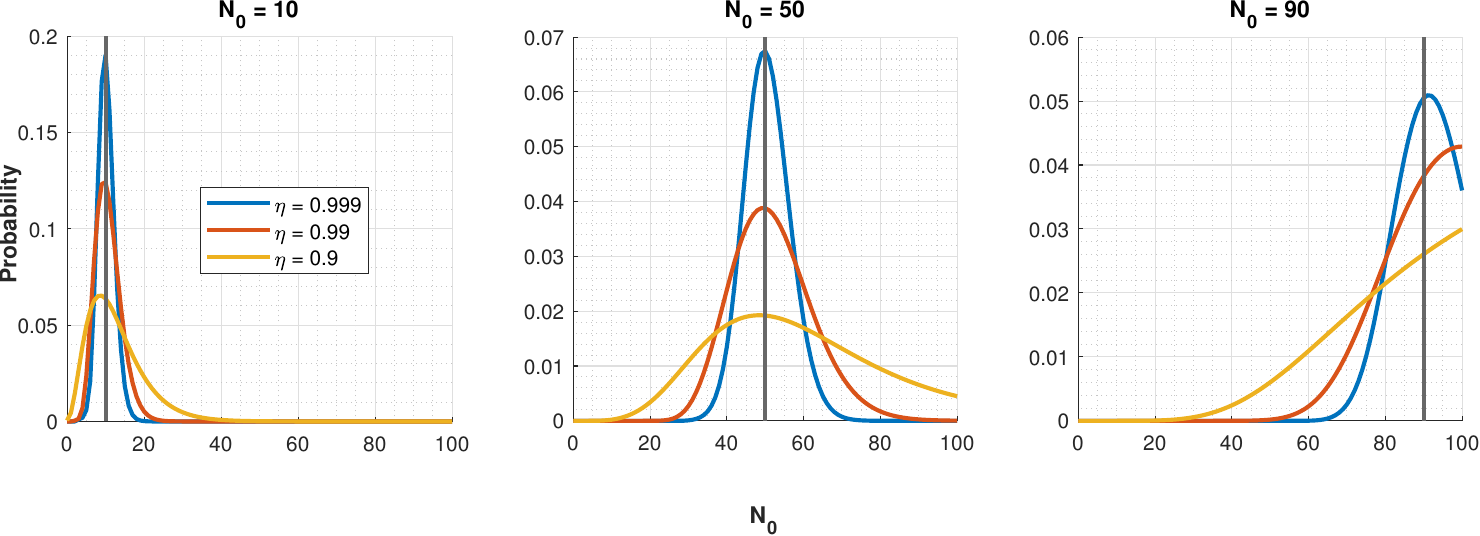}
	\caption{Posterior probability distributions for $N_0$, averaged over many trials, show that the estimate variance is highly sensitive to the loop efficiency $\eta$. The simulation performed $1000$ trials of the adaptive algorithm with parameters $\gamma = 0.9$ and $\nu = 1.0\times 10^{-6}$.}
    \label{fig:BarPlotBayesian}
\end{figure*}

Next, we compare this to a similar plot on the right side of Figure \ref{fig:EstimatePassiveAdaptive} for an adaptive trial. Similarly, the estimates fluctuate significantly in the beginning of the trial, before settling to a single value. One notable difference between the passive and adaptive setups is the effective length of the experiment. In the passive setup, the majority of the photons from the loop are detected within the first rounds, and thus most of the information about $N_0$ is acquired then as well. The remaining rounds are dedicated to measuring the last remaining photons. In the adaptive setup, in contrast, all the photons from the loop are detected before round $100$, allowing the measurement to be completed in fewer rounds. Despite outcoupling all the photons in a shorter number of rounds, the adaptive approach has lower outcoupling at the beginning so that the number of instances where multiple photons are outcoupled is reduced. The ability to yield an estimate of the initial photon number in a shorter time interval is important, since it reduces the number of photons that can be lost to loop loss before being detected. This shorter duration is also beneficial because it allows this architecture to perform PNR detection on a higher overall pulse rate.

Loop loss ($\eta$) is the single most significant contributor to uncertainty in the estimate of $N_0$. In Figure \ref{fig:BarPlotBayesian}, we have plotted the averaged Bayesian posterior distributions for $N_0$, for actual initial photon numbers of $N_0 = 10$, $N_0 = 50$ and $N_0 = 90$ (with $N_{max} = 100$), from simulations with different values of $\eta$. For $\eta = 0.999$, the uncertainty is quite small, and below the shot noise for all three values of $N_0$. For $\eta = 0.99$ and $\eta = 0.9$, however, the uncertainty in the estimate becomes larger, and it becomes more difficult to resolve the true value of $N_0$. 

\begin{figure}[btp] 
    \centering
	\includegraphics[width=1.0\linewidth]{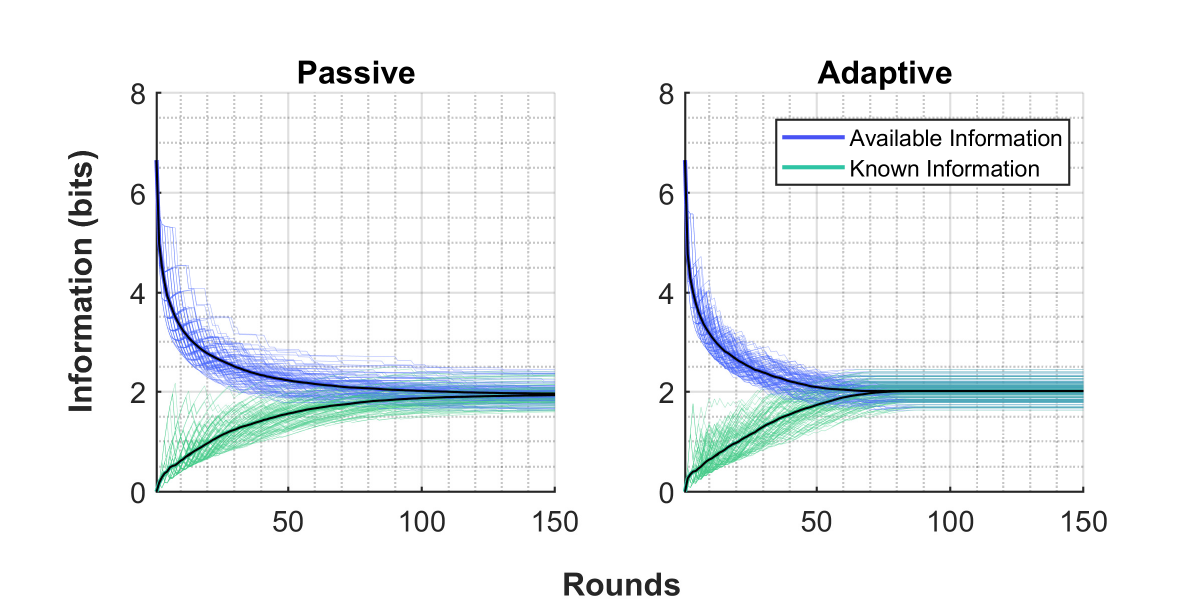}
	\caption{Comparison of passive-style ($\epsilon = 0.02$, left plot), and adaptive-style (right plot) setups illustrates how an adaptive strategy can extract the available information about $N_0$ more efficiently than a passive strategy. The known information (green lines) and available information (blue lines) for an ensemble of 100 trials are plotted in thin lines, while the means are shown with the weighted black lines. The simulations used initial photon number $N_0 = 40$, and parameters $\gamma = 0.9$, $\eta = 0.99$ and $\nu = 1.0\times 10^{-6}$.}
    \label{fig:InfoPassiveAdaptive}
\end{figure}

It is also important to keep track of the information gain from the detections and the remaining available information from the loop photons. These two quantities, as estimated during the simulation, are plotted in Figure \ref{fig:InfoPassiveAdaptive} for an ensemble of passive and adaptive trials, along with their mean values. We compared the adaptive strategy to the passive strategy with $\epsilon = 0.02$, which is the best-performing passive outcoupling rate for $N_0 = 40$. In both cases, there was initially $\log_2(N_{max})\approx 6.6$ bits of information available to gain from the photons in the loop, and $0$ bits of information gained. This is expected, since we are considering a uniform prior distribution, in which every value of $N_0$ is equally likely, and no detections have been made. Shortly after, the available information drops sharply to $\sim 3$ bits after $20$ rounds. Although only a few photons are lost in these initial rounds, this is enough to make close photon numbers effectively indistinguishable. For example, even if the expected number of lost photons is only one, the number of possible initial photons is increased by a factor of two, resulting in a loss of $\sim 1$ bit of information that was previously available. 

However, the known and available information subsequently evolve differently in both the passive and adaptive approach. While the two measures approach each other slowly in the passive case, as a result of the slow decay of the expected remaining loop photons, the convergence is rapid in the adaptive case, where the outcoupling is ramped up in the last rounds. Additionally, note that in the adaptive approach, the rate of information gain is approximately constant until the final photons leave the loop. This is a favorable feature for an adaptive approach to have, since it implies that information is gained more consistently and efficiently throughout the experiment.

\begin{figure*}[btp] 
    \includegraphics[width=1.0\linewidth]{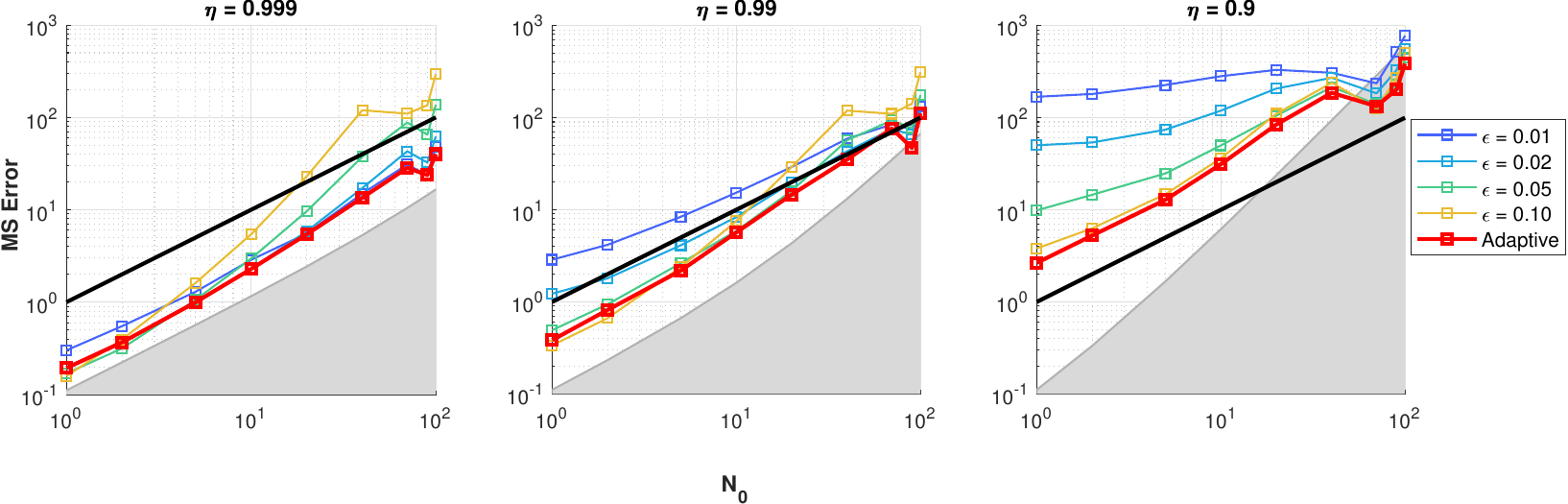}
	\caption{Comparison of the mean squared error (MSE) in the initial photon estimate $N_{est}$ between four passive approaches (constant outcoupling, thin green-blue lines) and the adaptive approach (thick red line), shows that the adaptive approach has a consistently lower error than each passive approach, over the full dynamic range from $1$ to $N_{max} = 100$. Performance is worse in every approach for lower loop efficiency $\eta$. The black line denotes the shot-noise limit $MSE = N_0$, while the shaded area denotes the theoretically optimal performance of an unbiased loop-based setup. The outcoupling probability $\epsilon$ took values from $0.01$ to $0.1$ for the passive approaches. The simulations used $1000$ trials, with parameters $\gamma = 0.9$ and $\nu = 1.0\times 10^{-6}$.}
    \label{fig:MSEPassiveAdaptiveMean}
\end{figure*}

\begin{figure*}[btp] 
    \includegraphics[width=1.0\linewidth]{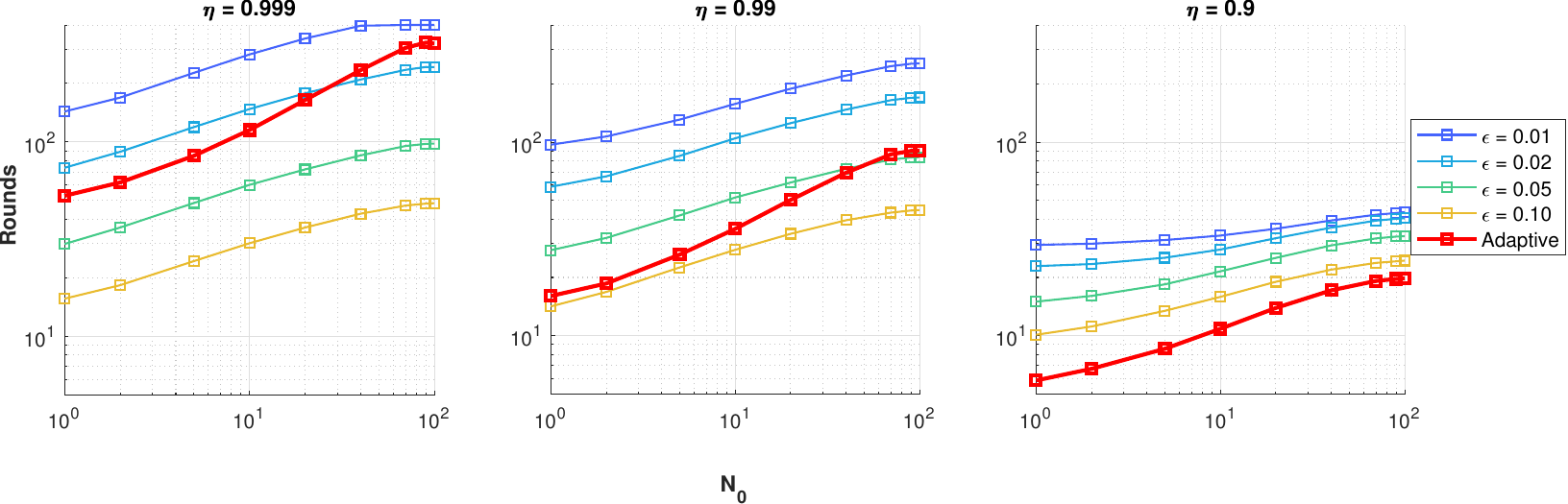}
	\caption{Comparison of the number of rounds needed for mean number of photons remaining in the loop (over many realizations) to fall below the threshold $N_t = 0.5$ between four passive approaches (constant outcoupling, thin green-blue lines) and the adaptive approach (thick red line), shows that while the adaptive approach does not consistently require a lower number of rounds than the passive approaches, it needs fewer rounds than passive approaches that achieve a similar MSE (see Figure \ref{fig:MSEPassiveAdaptiveMean}). Fewer rounds are required when the loop efficiency $\eta$ is lower, because more photons are lost to loop losses. For the passive approaches, the outcoupling probability $\epsilon$ took values from $0.01$ to $0.1$. The simulations used $1000$ trials, with parameters $\gamma = 0.9$ and $\nu = 1.0\times 10^{-6}$.}
    \label{fig:RoundsPassiveAdaptiveMean}
\end{figure*}

\begin{figure*}[btp] 
    \centering
    \includegraphics[width=1.0\linewidth]{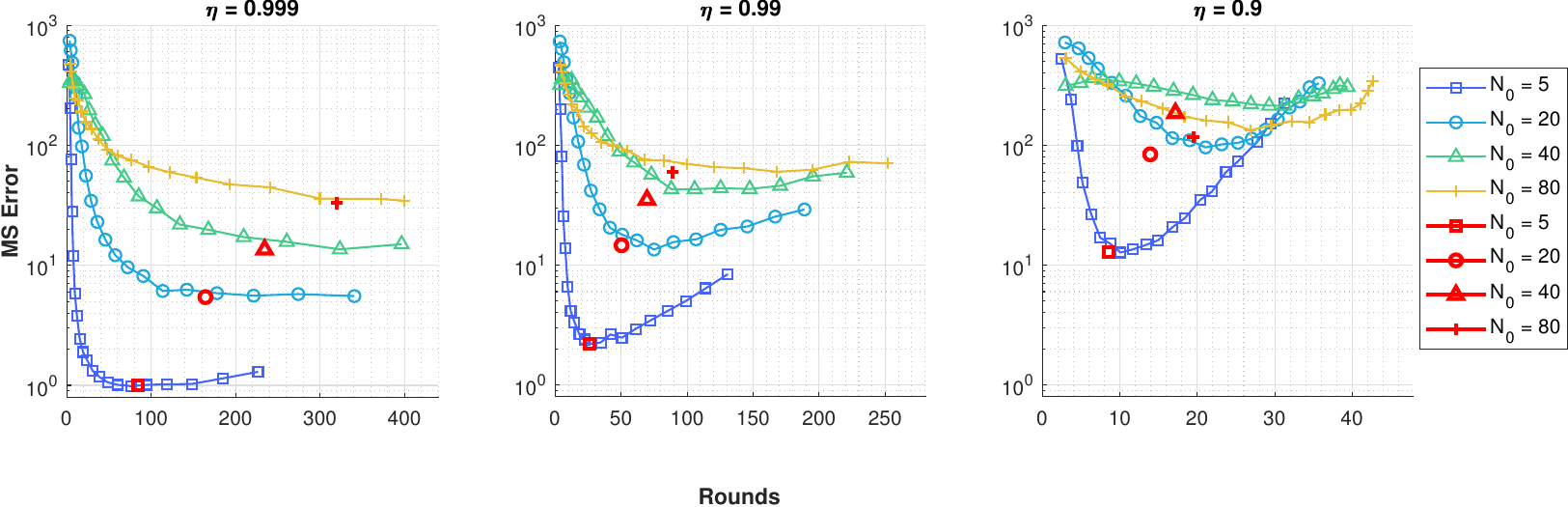}
    \caption{Comparison of the mean squared error and mean number of rounds needed to complete the measurement between the passive approaches (thin colored lines) and the adaptive approaches (bold red points) shows that the adaptive approach outperforms all corresponding passive approaches at each given value of $N_0$ and $\eta$. For the passive approaches, the outcoupling probability $\epsilon$ took values from $0.01$ to $1.0$. The simulations used $1000$ trials, with parameters $\gamma = 0.9$ and $\nu = 1.0\times 10^{-6}$.}
    \label{fig:MSErrorRoundsPassiveAdaptiveMean2}
\end{figure*}

In Figure \ref{fig:MSEPassiveAdaptiveMean}, we compare the MSE of the passive and adaptive methods for various values of $N_0$ and loop efficiency $\eta$. We can see that while the passive methods can perform well for some values of $\epsilon$ and $\eta$, only the adaptive approach performs well over a large dynamic range of $N_0$, while having steadily worsening performance as $\eta$ decreases. In the case of $\eta = 0.99$, the adaptive approach MSE remains below the shot noise for a dynamic range up to an order of magnitude higher than any of the passive approaches. 

One apparent problem of using the mean estimate of $N_0$ is its tendency to induce biases when $N_0$ is close to $0$ or $N_{max}$. For $N_0 \approx 0$, this is due to the extreme skew of the distribution, featuring a maximum for $N \approx N_0$ but a long tail for larger $N$, causing the estimate to have a positive bias. Conversely, for $N_0 \approx N_{max}$, the approximately symmetric distribution profile is cut off for $N > N_{max}$, causing the estimate to have a negative bias. To mitigate the estimator bias in a setup, one should choose $N_{max}$ carefully so that it is unlikely to create a cutoff effect. The MLE estimate does not suffer as acutely from these problems for smaller $N_0$, since it only depends on the distribution maximum, rather than the expectation value. However, it is also affected when $N_0 \approx N_{max}$, since the true maximum may be cut off in certain trials, causing a negative bias. 

The effect of the loop efficiency on the total number of rounds needed to complete the measurement is given in Figure \ref{fig:RoundsPassiveAdaptiveMean}. As expected, when the loop efficiency is higher, the photons can be stored for longer, resulting in more rounds (in some cases, more than $300$) needed to outcouple all photons. In contrast, for $\eta = 0.9$, neither the passive nor the adaptive approaches required more than $40$ rounds. From these plots, and comparing to the MS error plots in Figure \ref{fig:MSEPassiveAdaptiveMean}, we see that the adaptive approach often has the best combination of low error and fast measurement, within any value of $\eta$. An additional observation is that the number of rounds in the adaptive approach increases more rapidly than in a passive approach as a function of $N_0$. This is a consequence of the fact that in the adaptive approach, the average outcoupling decreases with $N_0$. 

Finally, we can combine the quantities of previous two plots into a single figure. In Figure \ref{fig:MSErrorRoundsPassiveAdaptiveMean2}, we compare the estimated MSE and mean number of rounds, between the adaptive approach (bold red points) and corresponding passive approaches (thin colored lines) for various values of $\eta$ and $N_0$. The passive approaches use a range of constant outcoupling values from $0.01$ to $1.0$. We see that the adaptive approach always outperforms the optimal passive approach with respect to this combination of parameters. Although the advantage is sometimes slight, this demonstrates the utility of the adaptive approach across a wide parameter range. The adaptive scheme is consistently as good as the optimal passive scheme in terms of MSE for $\eta$ close to $1$, as well as small photon number, while being $10-20\%$ faster to complete. However, for large photon number and high loop loss, the adaptive scheme outperforms any passive scheme in MSE, while being approximately two times faster than the optimal passive scheme. 

\section{Discussion}
\label{sec:discussion}

The adaptive loop-based PNR setup described here demonstrates performance advantages over alternative multiplexing-based architectures, with respect to dynamic range, accuracy, and speed. In particular, the results of the simulations over $1000$ trials show that introducing adaptive feedback into the loop-based setup improves the MSE of the estimate over a wide range of $N_0$. Although the passive approach with some choices of outcoupling rate $\epsilon$ can match the MSE of the adaptive approach over some $N_0$, they generally perform worse in another photon-number regime. For example, the $\eta = 0.99$ simulations in Figure \ref{fig:MSEPassiveAdaptiveMean} show that the $\epsilon = 0.1$ passive approach matches the adaptive approach at low photon numbers ($N_0 < 10$) but performs much worse at larger photon numbers. The opposite effect occurs for the $\epsilon = 0.01$ passive approach, which performs best at larger photon numbers. This demonstrates the ability of the adaptive algorithm to extend the dynamic range of the photon-loop PNR setup. 

The performance of the adaptive loop-based detector is close to the theoretically optimal performance for a loop-based detector, given a loop efficiency $\eta$ and detector quantum efficiency $\gamma$, as described in \hyperref[sec:methods]{Methods}. From Figure \ref{fig:MSEPassiveAdaptiveMean}, we see that for $\eta = 0.999$, the adaptive approach has mean-squared error only reaching up to twice as large as the theoretically optimal MSE for an unbiased estimator, over a wide range of $N_0$. For lower loop efficiencies such as $\eta = 0.9$, the MSE of the adaptive estimate is a full order of magnitude greater than the minimum MSE at low photon numbers. When loop losses are higher, we expect the dynamic range of a detector to be reduced, so this worse relative performance at low photon numbers is to be expected. Note that the MSE for both the adaptive and passive approaches dips below the theoretically optimal performance for large $N_0$. For $N_0$ close to the matrix dimension $N_{max}$, the posterior distribution exhibits a cutoff, which biases the estimate and allows the MSE to become artificially reduced. In contrast, the theoretically optimal estimate variance was calculated assuming the estimator is unbiased for all $N_0$. This artifact would disappear at these $N_0$ if $N_{max}$ is extended to some larger number.

The adaptive setup also demonstrates advantages in detection speed. Although it doesn't consistently complete detections in fewer rounds than all passive approaches, as seen in Figure \ref{fig:RoundsPassiveAdaptiveMean}, it needs fewer rounds than the passive approaches with similar performance. Figure \ref{fig:MSErrorRoundsPassiveAdaptiveMean2} illustrates this trade-off in a more compact way, and shows that the adaptive approach outperforms the best passive approach in this combination of quantities, over a wide dynamic range. This is a significant advantage of the adaptive approach since detection rate is often important for PNR detectors.

The adaptive loop-based detector also compares favorably to other PNR detector architectures that use multiplexing. For small photon numbers, the performance is essentially limited by the detector efficiency $\gamma$, no matter which architecture is used, as long as the number of detector bins is sufficiently larger than $N_0$. However, the performance at higher $N_0$ depends greatly on the details of the setup. For beam splitter-based spatial multiplexing detectors, or time multiplexing detectors using beam splitters and delay lines, the performance becomes limited by the number of detector bins into which the signal can be multiplexed. In fact, to maintain a given photon number resolution, the number of detector bins must scale quadratically with $N_0$ \cite{jonssonEvaluatingPerformancePhotonnumberresolving2019}. In these setups, the number of detector bins places a hard limit on the dynamic range. In contrast, the dynamic range of a loop-based detector is essentially limited by the loop efficiency $\eta$, with estimate error increasing quadratically with $N_0$ for $N_0 \gtrsim (1 - \eta)^{-1}$. 

However, there are challenges associated with implementing the adaptive approach in practice. Firstly, the process of computing the optimal outcoupling $\epsilon_k$ for each round is fairly computationally demanding as implemented here. Beginning from the probability matrix $\mathbf{P}(\vec{d}_k)$, which is a square matrix of dimension $N_{max} + 1$, the expected information gain must be calculated for many potential values of $\epsilon_k$. This involves computing multiple matrix multiplications and many logarithms each round, and must be computed fast enough to update the outcoupling in time for the next round. This would require a dedicated fast computing module such as a field programmable gate array (FPGA) to store the probability matrix information, as well as receive the detection results, perform the matrix calculations, and send the updated outcoupling value to the setup in a very short time on the order of microseconds or hundreds of nanoseconds. 

Alternatively, it might be useful to find a heuristic to determine the outcoupling that does not involve such extensive calculations. One particularly economical method could involve updating the outcoupling solely based on the current outcoupling value and the detection results, for example by incrementing or decrementing the outcoupling by a fixed percentage depending on whether a click was recorded or not. In such an approach, all the photon number estimation calculations would be performed after the fact. A slightly more sophisticated technique would involve performing at least some of this estimation in real time, but using a simpler algorithm to find an optimal outcoupling $\epsilon_k$. Each of these approaches would have their own tradeoffs in speed, computational resources, and accuracy, which should be evaluated with respect to the particular loop-based detector implementation. 

In addition to its good performance, the adaptive algorithm considered here has the advantage of having a simple interpretation, as well as being suitable for any given prior photon number distribution. However, since it employs a greedy local optimization of information gain to information loss in the next round (see Equation \eqref{eq:InfoRatio}), it is likely not globally optimal. In particular, it ignores the fact that most of the information lost in the initial rounds of the experiment is unavoidable, regardless of the chosen outcoupling values. An algorithm that accounts for how the outcoupling rate in a given detection round affects the global information gain would likely have performance closer to the theoretical limit. However, this is more difficult to achieve in practice, since one would need to estimate the total expected information gain during each detection round, which is much more computationally demanding. Alternatively, one could use an algorithm that attempts to maximize the expected number of detector clicks in the experiment, but this would also likely be significantly more computationally demanding than a greedy optimization of information gain. 

The most significant challenges associated with experimentally implementing this setup are the low-loss fast switch and the low-loss storage line. As seen from the simulation results, the uncertainty in the results is very sensitive to the loop efficiency $\eta$. Since every photon passes through the coupler on each pass, the loss associated with the coupler is also a component of the loop efficiency. Thus, necessary for implementation are both a low-loss coupler that can respond in a time $\tau$ that allows for adaptive feedback, and a storage fiber that has low loss over a length $c\tau/n$, where $n$ is the refractive index of the fiber.

There are several options for implementing this setup. A free-space setup could be implemented using an EOM as the fast switch, with free-space fiber couplers and a fiber providing the delay loop. The typical loss in a low-loss fiber is approximately $0.2 \un{dB/km}$, so most of the loss in the delay loop would result from insertion loss or losses associated with the fiber coupler. Alternatively, a free-space delay line could be implemented using an optical cavity loop, together with an EOM functioning as the fast switch. This has been demonstrated using a polarizing beam splitter and Pockels cell with delay time $\tau = 10\un{ns}$ and loop efficiency $\eta = 0.988$ \cite{kanedaHighefficiencySinglephotonGeneration2019}. 

All-fiber options are also important to consider. By outcoupling the fiber loop into an in-fiber EOM and coupler, the outcoupling between fiber and free-space could be minimized.  However, insertion losses between the fibers could still significantly contribute to loop losses. An all-integrated approach, in which the EOM, coupler, and fiber loop are integrated and manufactured on a chip, could help to minimize the coupling losses between each component of the detector, but at the disadvantage of short delay times and high on-chip losses, which would result in a relatively large loop loss. 

\section{Conclusion}
\label{sec:conclusion}

In this paper, we have introduced a method to perform adaptive single-shot measurements of photon number using a storage loop. The algorithm was designed to choose the loop outcoupling that optimizes the ratio between the expected information gained due to detections and the expected information lost due to losses, based on the sequence of recorded detections. Through simulations of the loop setup ($N_{trials} = 1000$) over a range of parameters and initial photon numbers, we found that the adaptive approach achieved improvements in accuracy, dynamic range, and speed relative to any passive setup with constant outcoupling. 

In particular, the mean squared error (MSE) of the adaptive approach's estimate was equal to or less than that of any passive approach, over a wide range of photon number from $0$ to $100$. No individual choice of outcoupling rate $\epsilon$ was able to match the performance of the adaptive algorithm over this whole range. This demonstrates that improvements in accuracy and dynamic range possible with an adaptive approach. Furthermore, in comparison to the passive approaches that achieve a similar MSE, the adaptive approach required fewer rounds to complete the detection. Since detection rate is often an important quality for PNR detectors, this is also a significant advantage of the adaptive approach.

Despite the advantages, there are several limitations to address in this adaptive loop-based PNR detector. Since both low loop loss and rapid feedback are vital to the detector's performance, the most important limitation for experimental implementation is the difficulty of designing or obtaining a rapid low loss variable coupler. Additionally, the adaptive algorithm for choosing the loop outcoupling rate is likely too slow and computationally intensive for the needed real-time adjustments. The use of a cutoff $N_{max}$, which is required for the Bayesian estimation calculations, also significantly affects the estimate when the photon number is close to this cutoff. 

With these limitations in mind, there are several possible extensions of this work that may be worthwhile to investigate. For instance, although the adaptive algorithm used here shows performance advantages over the passive approaches, it is unlikely to be the optimal approach. It would be interesting to determine if there is an approach that has better performance, or performs as well with fewer computations, than the one used here. Another possible extension to this work is analyzing the performance of such a setup for a detector that already has some photon-number resolution. This could be accomplished either using another multiplexed setup, or a detector that can resolve multi-photon events such as an SNSPD or TES. By providing more information each round, the detector has an overall performance that is less dependent on the loop loss. As a result, loop-based multiplexing can also be a simple way to extend the dynamic range of PNR detectors while maintaining their high resolution. 

In many applications, one has some prior knowledge of the photon number distribution, which can be used to improve the single-shot estimation of $N_0$ and optimize the outcoupling rate by acting as a non-uniform prior distribution for the adaptive algorithm. For example, if we know that the photon number distribution is Poissonian, as occurs in a coherent state, the initial outcoupling rate can be optimized to maximize the expected information gained for the known mean photon number. Alternatively, if we know that the number of incident photons will likely be one of two discrete values $N_1$ and $N_2$, then the outcoupling rates can be optimized by the adaptive algorithm to best distinguish between these two cases. 

The loop-based PNR detector with adaptive feedback presented in this paper demonstrates improved performance capabilities in comparison to other passive approaches. The flexibility of the paradigm, illustrated by the potential extensions, may allow adaptive photon storage loops to be useful for many diverse applications in optical quantum computing and metrology. 

\bibliography{PhotonLoopPaper}
\bibliographystyle{unsrt}

\section{Acknowledgements}

We thank the NSERC USRA program for supporting this work. B.B. acknowledges the support of the Banting postdoctoral fellowship. RWB acknowledges support through the Natural Sciences and Engineering Research Council of Canada, the Canada Research Chairs program, the Canada First Research Excellence Fund award on Transformative Quantum Technologies, US National Science Foundation Award 2138174, and US Department of Energy award FWP 76295.

\appendix

\section{Bayesian Update Probability Calculation}
\label{sec:appendixA}

Calculating the probability $P(N_{k}, d_{k}| N_{k-1})$ involves multiple steps. First, we calculate the probability that $N_k$ photons remain, accounting for both loop efficiency ($\eta$) and the outcoupling rate ($\epsilon_k$). This probability is given by a binomial distribution:

\begin{equation}
    \begin{split}
        P(N_{k} | N_{k-1}) 
        & = B(N_k; N_{k-1}, \eta(1-\epsilon_k)),
    \end{split}
\end{equation}

where $B(k; n, p)$ is the binomial distribution function:
\begin{equation}
    \begin{split}
        B(k; n, p) & = \binom{n}{k} p^k (1-p)^{n-k}.
    \end{split}
\end{equation}

However, to factor in the detection results, we need to find the probability that $N_k$ photons remain in the loop, and that $m$ photons are outcoupled from the loop to the detector. This probability is given:

\begin{equation}
    \begin{split}
        P(N_{k}, m | N_{k-1}) 
        & = B(N_k+m; N_{k-1}, \eta) \\
        & \times B(m; N_k+m, \epsilon_k).
    \end{split}
\end{equation}

Given that $m$ photons are sent to the detector, the probability that $d_k = 0$ (i.e. no photons detected), is the probability that no dark counts occur and none of the $m$ photons trigger the detector. Here, we are additionally assuming that each of the $m$ photons interact with the detector independently of any others, and no nonlinear effects occur. Under these conditions, this probability is:
\begin{equation}
    \begin{split}
        P(d_k=0 | m) 
        & = (1-\nu)(1-\gamma)^m.
    \end{split}
\end{equation}

Several useful binomial distribution identities include its behavior under convolution:
\begin{equation}
    \begin{split}
        \sum_{k=m}^{n} B(k; n, p) B(m; k, q) & = B(m; n, pq),
    \end{split}
\end{equation}
its behavior under multiplication by $q^m$:
\begin{equation}
    \begin{split}
        q^m B(m; n, p) & = \left[ \frac{1-p}{1-pq}\right]^{n-m} B(m; n, pq),
    \end{split}
\end{equation}
and its behavior under transformation of probability $p \to 1 - p$:
\begin{equation}
    \begin{split}
        B(m; n, p) & = B(n-m; n, 1-p).
    \end{split}
\end{equation}

Using these identities, we find that the total probability that $d_k=0$ is:
\begin{equation}
    \begin{split}
        & \rho(N_k; N_{k-1}, \eta, \epsilon_k, \gamma, \nu) \\
        & = \sum_{m=0}^{N_{k-1}-N_k} P(d_k=0 | m) P(N_{k}, m | N_{k-1}) \\
        & = \sum_{m=0}^{N_{k-1}-N_k} (1-\nu)(1-\gamma)^m \\
        & \times B(N_k+m; N_{k-1}, \eta) B(m; N_k+m, \epsilon_k).
    \end{split}
\end{equation}

This simplifies to:
\begin{widetext}
\begin{equation}
    \begin{split}
        \rho(N_k; N_{k-1}, \eta, \epsilon_k, \gamma, \nu) & = \sum_{m=0}^{N_{k-1}-N_k} (1-\nu)\left[\frac{1-\epsilon_{k}}{1-\epsilon_k(1-\gamma)}\right]^{N_k} B(N_k+m; N_{k-1}, \eta) B(m; N_k+m, \epsilon_k(1-\gamma))\\
        & = \sum_{m=0}^{N_{k-1}-N_k} (1-\nu)\left[\frac{1-\epsilon_{k}}{1-\epsilon_k(1-\gamma)}\right]^{N_k}  B(N_k+m; N_{k-1}, \eta) B(N_k; N_k+m, 1-\epsilon_k(1-\gamma))\\
        & = (1-\nu)\left[\frac{1-\epsilon_{k}}{1-\epsilon_k(1-\gamma)}\right]^{N_k} B(N_k; N_{k-1}, \eta(1-\epsilon_k(1-\gamma)))\\
        & = (1-\nu)(1 - \eta\epsilon_k\gamma)^{N_{k-1}} B\left(N_k; N_{k-1}, \frac{\eta(1-\epsilon_k)}{1 - \eta\epsilon_k\gamma}\right).
    \end{split}
\end{equation}

Thus, the total probability for both the $d_k = 0$ and $d_k = 1$ cases is given by:
\begin{equation}
    \begin{split}
        P(N_{k}, d_{k}| N_{k-1}) & = 
        \begin{cases}
            \rho(N_k; N_{k-1}, \eta, \epsilon_k, \gamma, \nu) & d_{k} = 0\\
            B(N_k; N_{k-1}, \eta(1-\epsilon_k)) - \rho(N_k; N_{k-1}, \eta, \epsilon_k, \gamma, \nu) & d_{k} = 1\\
        \end{cases},
    \end{split}
\end{equation}
where
\begin{equation}
    \begin{split}
        \rho(N_k; N_{k-1}, \eta, \epsilon_k, \gamma, \nu) & = (1-\nu)(1 - \eta\epsilon_k\gamma)^{N_{k-1}} B\left(N_k; N_{k-1}, \frac{\eta(1-\epsilon_k)}{1 - \eta\epsilon_k\gamma}\right).
    \end{split}
\end{equation}
\end{widetext}

\section{Variables and Symbols}
\label{sec:appendixB}

\begin{table*}[btp] 
    \begin{ruledtabular}
        \begin{tabular}{|c|p{0.7\textwidth}|}
            Symbol & Explanation\\
            \hline
            $N_0$ & Number of photons initially in the loop.\\
            $N_k$ & Number of photons in the loop after round $k$.\\
            $\eta$ & Loop efficiency\\
            $\gamma$ & Detector efficiency\\
            $\nu$ & Detector dark count rate\\
            $\epsilon$ & Loop outcoupling rate\\
            $\epsilon_k$ & Loop outcoupling rate in round $k$ (for adaptive setup)\\
        \end{tabular}
    \end{ruledtabular}
    \caption{List of all variables and symbols associated with the loop-based PNR detector setup in this paper.}
    \label{table:symbols1}
\end{table*}

\begin{table*}[btp] 
    \begin{ruledtabular}
        \begin{tabular}{|c|p{0.7\textwidth}|}
            Symbol & Explanation\\
            \hline
            $d_k$ & Detection result in round $k$. If $d_k = 0$, no click was recorded, and if $d_k = 1$, a click was recorded.\\
            $\vec{d}_k$ & Series of detection results up to round $k$.\\
            $P(N_k, N_0 | \vec{d}_k)$ & Posterior probability of $N_k$ photons being in the loop during round $k$, and $N_0$ photons in the loop initially, given the detection results $\vec{d}_k$.\\
            $P(N_k, \vec{d}_k | N_0)$ & Probability of yielding $N_k$ photons in the loop during round $k$, and the detection results $\vec{d}_k$, given $N_0$ photons initially in the loop.\\
            $P(N_0)$ & Prior probability of $N_0$ photons initially being in the loop.\\
            $P(N_k, d_k | N_{k-1})$ & Probability of yielding $N_k$ photons in the loop during round $k$, and the detection result $d_k$, given $N_{k-1}$ photons in the loop in the previous round.\\
            $\rho(N_k; N_{k-1})$ & Probability of yielding $N_k$ photons in the loop during round $k$, and a `no click' detection result, given $N_{k-1}$ photons in the loop in the previous round.\\
            $B(k;n,p)$ & Binomial distribution function with probability $p$, choosing $k$ out of $n$ objects.\\
            $\mathbf{P}(\vec{d}_k)$ & Matrix representation of $P(N_k, \vec{d}_k | N_0)$.\\
            $\mathbf{R}(d_k)$ & Matrix representation of $P(N_k, d_k | N_{k-1})$.\\
            $D_{KL}(P||Q)$ & Kullback-Leibler (KL) divergence of the probability distribution $P$ against the probability distribution $Q$. Usually, $P$ is taken to be the `true' distribution, and $Q$ the model.\\
            $I_{G,k}$ & Estimated information gained from experimental detection results $\vec{d}_k$ as of round $k$.\\
            $I_{A,k}$ & Estimated information available to be learned based on the photons remaining in the loop, as of round $k$.\\
        \end{tabular}
    \end{ruledtabular}
    \caption{List of all variables and symbols used to simulate the passive and adaptive approaches in this paper.}
    \label{table:symbols2}
\end{table*}

\begin{table*}[btp] 
    \begin{ruledtabular}
        \begin{tabular}{|c|p{0.7\textwidth}|}
            Symbol & Explanation\\
            \hline
            $N_{est}(\vec{d}_k)$ & Estimated initial number of photons based on detection results $\vec{d}_k$.\\
            $\mathrm{Var}_{est}(\vec{d}_k)$ & Estimated variance in the initial number of photons based on detection results $\vec{d}_k$.\\
            $N_{MLE}(\vec{d}_k)$ & Maximum likelihood estimate (MLE) of the initial number of photons based on detection results $\vec{d}_k$.\\
            $\langle \cdot \rangle$ & Average of some quantity over many trials.\\
            $\mathrm{Var}(N_{est})$ & Variance of $N_{est}(\vec{d}_k)$ over many trials.\\
            MSE & Mean square error $N_{est}(\vec{d}_k)$ relative to $N_0$, over many trials.\\
        \end{tabular}
    \end{ruledtabular}
    \caption{List of all variables and symbols used to compare the passive and adaptive approaches in this paper.}
    \label{table:symbols3}
\end{table*}

There are many variables and symbols used in this paper, so the purpose of this appendix is to enumerate the meanings of all of them. Table \ref{table:symbols1} lists the basic parameters associated with the loop-based PNR detector setup, as discussed in the \hyperref[sec:intro]{Introduction}. Table \ref{table:symbols2} explains the variables and distributions relevant for the simulation and Bayesian inference, as discussed in the \hyperref[sec:mathback]{Mathematical Background} section. Finally, Table \ref{table:symbols3} lists the variables used to analyze and compare the various passive and adaptive approaches, as discussed in the \hyperref[sec:methods]{Methods} and \hyperref[sec:results]{Results} sections.

\end{document}